\documentclass[]{aa}
\usepackage{graphics,psfig}

\begin{document}

\title{The complete ISO spectrum of NGC~6302\thanks{Based on
observations with ISO, an ESA project with instruments
funded by ESA Member States (especially the PI countries: France,
Germany, the Netherlands and the United Kingdom) and with the
participation of ISAS and NASA}}
\author{F.J. Molster\inst{1},
T.L. Lim\inst{2},
R.J. Sylvester\inst{3},
L.B.F.M. Waters\inst{1,4},
M.J. Barlow\inst{3},
D.A. Beintema\inst{5},
M. Cohen\inst{6},
P. Cox\inst{7} and
B. Schmitt\inst{8}
}
\institute{
Astronomical Institute 'Anton Pannekoek', University of Amsterdam,
Kruislaan 403, 1098 SJ Amsterdam, The Netherlands
\and
Space Science Department, Rutherford Appleton Laboratory, Chilton, Didcot,
OX11 0QX, United Kingdom
\and
Department of Physics and Astronomy, University College London, Gower Street,
London WC1E 6BT, United Kingdom
\and
Instituut voor Sterrenkunde, K.U. Leuven, Celestijnenlaan 200B,
3001 Heverlee, Belgium
\and
SRON Space Research Laboratory, P.O. Box 800, NL-9700 AV Groningen, The Netherlands
\and
Radio Astronomy Laboratory, 601 Campbell Hall, University of California,
Berkeley, CA 94720
\and
Institut d'Astrophysique Spatiale, B\^{a}t 121, Universit\'{e} de Paris XI, 
F-91405 Orsay Cedex, France
\and
Laboratoire de Planetologie de Grenoble, Universite J. Fourier - CNRS,
Batiment D de Physique, BP 53, 38041 Grenoble Cedex 9, France}

\offprints{F.J. Molster: fjmolster@mvainc.com}\mail{F.J. Molster,
MVA Inc, Oakbrook Parkway 5500, suite 200, Norcross, GA 30093,
USA}

\date{received\dots, accepted\dots}

\authorrunning{F.J. Molster et al.}
\titlerunning{The complete ISO spectrum of NGC~6302}

\abstract{
We present the combined Infrared Space Observatory Short-Wavelength
Spectrometer and Long-Wavelength Spectrometer 2.4--197~$\mu$m spectrum of the
Planetary Nebula NGC~6302 which contains in addition to strong atomic lines,
a series of emission features due to solid state components. The
broad wavelength coverage enables us to more accurately  identify and determine
the properties of both oxygen- and carbon-rich circumstellar dust. A simple
model fit was made to determine the abundance and typical temperature
of the amorphous silicates, enstatite and forsterite.
Forsterite and enstatite do have roughly the same abundance and temperature.
The origin and
location of the dust in a toroidal disk around the central star are discussed.}

\maketitle


\section{Introduction}

The bipolar planetary nebula (PN) NGC~6302 is one of the brightest PN in our
galaxy, and its central star is one of the hottest stars known 
(Ashley \& Hyland 1988; Pottasch et al. 1996). 
These properties suggest that the progenitor of NGC~6302
was massive, which is consistent with the high nebular abundance of nitrogen
(Aller et al. 1981). The morphology of the nebula is strongly bipolar, with a
dense, dusty torus  (Lester \& Dinerstein 1984) and 
ionized polar regions as traced by the free-free radio
emission (Gomez et al. 1993). Roche \& Aitken (1986) detected 8.6
and 11.3~$\mu$m  polycyclic aromatic hydrocarbon (PAH) band emission from 
NGC~6302, usually associated with carbon-rich material, 
while the presence of oxygen-rich material was indicated by the detection
of an OH-maser (Payne, Phillips \& Terzian 1988) and an 18~$\mu$m
amorphous silicate emission feature (Justtanont et al. 1992).
NGC~6302 therefore can be
considered as an important test of stellar evolution models: the nebular
abundances provide constraints on the chemical evolution and nucleosynthesis
processes, and the thermal emission from dust gives information on the mass
loss and its geometry at the end of the AGB phase.

The infrared brightness and unique properties of NGC~6302 made it an excellent
target for study with the Short Wavelength (SWS; de Graauw et al. 1996) and
Long Wavelength (LWS; Clegg et al. 1996) Spectrometers on board the Infrared
Space Observatory (ISO; Kessler et al. 1996). Several papers have already
reported on some aspects of these ISO observations. One of the most remarkable
results of the ISO observations of NGC~6302 was the discovery of strong
emission due to crystalline silicates (Waters et al. 1996; Barlow 1998;
Beintema 1998), as well as several other solid state emission bands, including
those of water ice.
A full inventory and characterization of these emission bands requires access
to the entire 2.4--197~$\mu$m wavelength range that ISO offers.
In this paper, we
present for the first time the full grating spectrum of NGC~6302.
The present analysis of the spectrum will concentrate on the dust features.

This paper is organized as follows. In Sect.~2 we briefly describe the
observations and data reduction. Sect.~3 discusses the placement of the
continuum and presents the continuum subtracted spectrum. We identify most of the
solid state bands. In Sect.~4 the results of
a simple model to fit the spectral dust features and the continuum 
are presented and in Sect.~5 we discuss the
results of our analysis.

\section{The observations and data reduction}

The SWS spectrum from 2.4 to 45 $\mu$m of NGC~6302 was obtained with ISO on
February 19, 1996 as an AOT01, with the highest resolution (speed 4). 
The total integration time was 6528 seconds.
The spectra were reduced using the SWS offline processing
software, version 7.0. For a description of flux and wavelength calibration
procedures, see Schaeidt et al. (1996) and Valentijn et al. (1996). The
main fringes in the 12.0 -- 29.5 $\mu$m part of the spectrum  were removed
using the Interactive (IA) data reduction package routine {\sc FRINGES}. Major
irregularities due to glitches and large drops or jumps in each detector were
removed by hand. A comparison between the different detectors scanning the
same wavelength region was used to determine the location of such jumps. The
12 SWS sub-spectra, when combined into a single spectrum, can  show jumps in
flux levels at band edges due to imperfect flux calibration or dark current
subtraction. We have adjusted the different sub-bands, 
according to the expected
source of discrepancy in each band, to form a continuous spectrum. 
Note that all the adjustments are well within the photometric absolute 
calibration uncertainties of 20\% (Schaeidt et al. 1996). 
The relative error depends on band and flux but is typically much less 
than 5\%.
The final resolving power ($\lambda / \Delta\lambda $) is 1000, apart from
band~2C (7 -- 12 $\mu$m) and 3E (27.5 -- 29 $\mu$m) which have a resolving 
power of 500 and from band~1A (2.4 -- 2.6 $\mu$m) which was rebinned to a 
resolving power of 2000.

\begin{figure}[ht]
\centerline{\psfig{figure=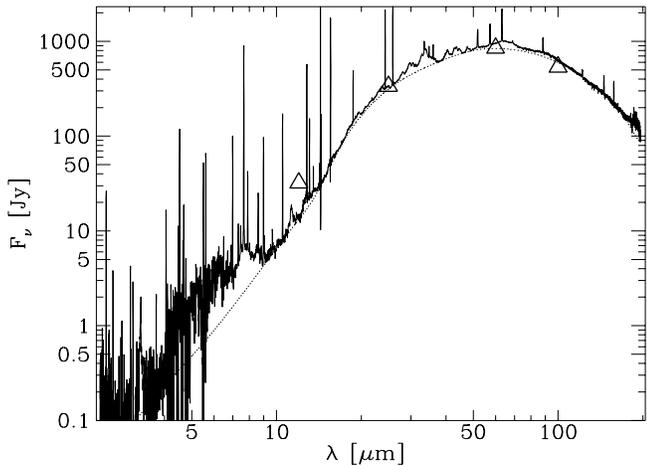,width=85mm,angle=270}}
\caption[]{The combined SWS and LWS spectrum of NGC~6302 (solid line).
The IRAS broad-band flux densities are shown as triangles. Note that the
IRAS~12~$\mu$m band is severely influenced by a series
of strong atomic lines. The dashed line is
the spline continuum fit used for the continuum subtraction 
(see text and Fig~\ref{fig:contsub}).}
\label{fig:spec}
\end{figure}

The LWS spectrum from 40 to 197 $\mu$m was obtained by combining seven full-range
grating scan AOT L01 observations, obtained between 1996 August 31 and 1997
September 23, in revolutions 289, 482, 489, 503, 510, 671, and 678. An
off-source spectrum, of a position 8 arcmin from NGC~6302, was also obtained in
revolution 289. For a description of the flux and wavelength calibration
procedures see Swinyard et al. (1996). Each of the seven observations consisted
of six fast grating scans, with 0.5~s integration at each commanded grating
position. The observations taken in revolutions 289 and 671 were sampled at 1/4
of a spectral resolution element, the latter being 0.3~$\mu$m in second order
(detectors SW1--SW5 $\lambda\leq$93$\mu$m) and 0.6~$\mu$m in first order
(detectors LW1--LW5 $\lambda\geq$80$\mu$m). All other observations were made
with a sampling of 1/8 of a resolution element. The total on-target integration
time was 13243~s. 
The data were reduced with the LWS offline processing software
version 7.0. No scaling between the datasets was necessary before combining
them to form a single dataset. The sub-spectra from the ten LWS detectors were
individually rescaled to give the same flux in the regions of overlap, adopting
a polynomial fit through all detectors to give scaling values. Each detector
was scaled by a factor within 10\% of unity. This is consistent with the 
formal absolute error of 10\% per detector for a medium flux
point source such as NGC~6302 (Swinyard et al. 1996). The scans were then
averaged giving a relative error of no more than 3\% per point. The ten
sub-spectra were then merged to form a complete spectrum from 40 to 197$\mu$m. 

The whole SWS spectrum has to be multiplied by 1.22 to match the LWS-spectrum.
The LWS observation was taken at the central position but for the SWS
observation the satellite was pointed 5 arcsec away from the centre of
NGC~6302. It should be noted that the sensitivity of the SWS decreases 
with distance from the centre.
Another factor in the photometric differences between the two ISO
spectrometers is the aperture size. The aperture of LWS is circular
with a diameter of about 80 arcsec and the aperture size for SWS is rectangular
and varies from band to band, from 14 arcsec $\times$ 20 arcsec for the 
shortest wavelengths to 20 arcsec $\times$ 33 arcsec for the longest 
wavelengths. The final spectrum is shown in
Fig.~\ref{fig:spec}, together with the IRAS broadband photo\-metry points. The
differences found between the IRAS fluxes and the ISO-fluxes convolved
with the IRAS system response curves are well within the quoted IRAS,
SWS and LWS flux uncertainties. The IRAS 12 micron
flux, which seems to deviate, is actually consistent with our ISO observations
when the forbidden emission lines are taken into account.
The combined spectrum shows a wealth of atomic
lines which are described in Beintema \& Pottasch (1999) and a series of broad
emmission bands due to solid state features, which will be described in this 
paper.

\section{The dust features}
\label{sec:dust}

\begin{figure*}[ht]
\resizebox{\hsize}{!}{\includegraphics{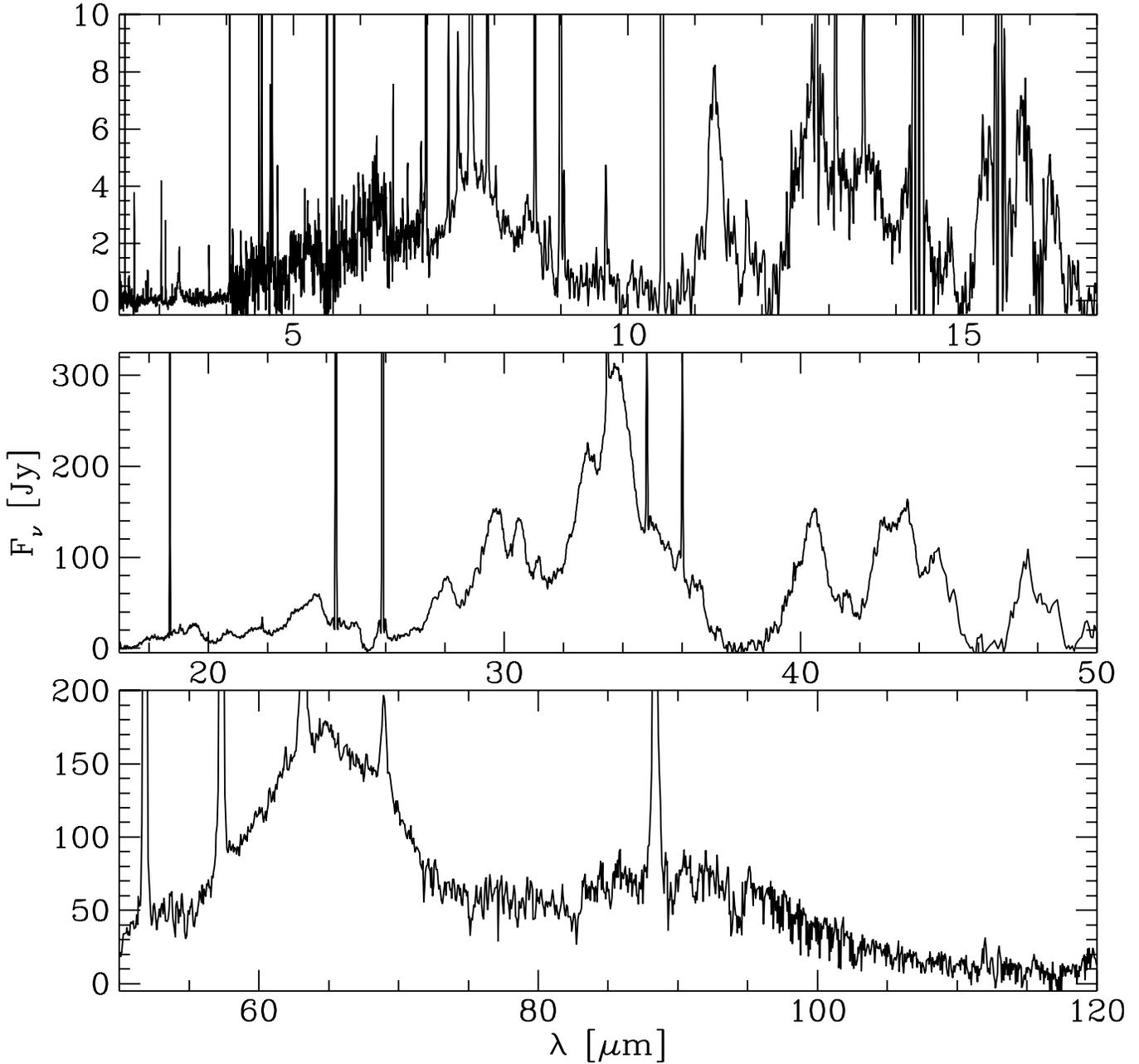}}
\caption[]{
The continuum-subtracted spectrum of NGC~6302 from 2.4
to 120~$\rm \mu m$. The region from 2.4 to 17~$\rm \mu m$
(top panel) is characterized by the presence of C-rich dust 
features. At longer wavelengths, the spectrum becomes
dominated by O-rich dust with the crystalline 
silicates bands in the 17 to 50~$\rm \mu m$ region 
(middle panel) and broad structures of crystalline
$\rm H_2O$ ice and probably hydrous silicates 
beyond 50~$\rm \mu m$ (bottom panel).}
\label{fig:contsub}
\end{figure*}

\begin{table*}
\caption[]{The characteristics of the dust features in the spectrum of
NGC~6302. The quoted errors are derived by multiple independent fits to
the different spectral features.}
{\small
\begin{tabular}{llllllllll}
\hline

\multicolumn{1}{c}{$\lambda$} &
\multicolumn{1}{c}{FWHM} &
\multicolumn{1}{c}{$I_{\mbox{peak}}/I_{\mbox{cont}}$} &
\multicolumn{1}{c}{Band flux} &
Identification  \\
\multicolumn{1}{c}{$ \mu$m}    &
\multicolumn{1}{c}{$ \mu$m}    &
 &
\multicolumn{1}{c}{$10^{-13}$ Wm$^{-2}$}     &
 & \\

\hline
 3.29 (0.01) & 0.05 (0.01) & 6.3   (2.7)   & 0.143 (0.009) & PAH C-H stretch \\
 3.41 (0.01) & 0.05 (0.02) & 2.1   (0.7)   & 0.027 (0.006) & PAH C-H stretch \\
 6.24 (0.01) & 0.20 (0.01) & 1.71  (0.04)  & 0.341 (0.010) & PAH C-C stretch \\
 7.66 (0.05) & 0.69 (0.02) & 1.74  (0.08)  & 0.87  (0.26)  & PAH C-C stretch \\
 8.56 (0.04) & 0.36 (0.03) & 1.42  (0.08)  & 0.239 (0.012) & PAH C-H bending in plane \\
11.01 (0.03) & 0.13 (0.07) & 1.180 (0.029) & 0.05  (0.04)  & PAH+artifact?\\
11.29 (0.01) & 0.27 (0.07) & 1.77  (0.07)  & 0.49  (0.11)  & PAH C-H bending out of plane \\
18.10 (0.02) & 0.75 (0.13) & 1.125 (0.018) & 0.96  (0.17)  & enstatite + amorphous silicate \\
18.97 (0.01) & 0.68 (0.06) & 1.128 (0.006) & 1.05  (0.07)  & unidentified + amorphous silicate? \\
19.56 (0.01) & 0.45 (0.02) & 1.135 (0.009) & 0.79  (0.09)  & forsterite \\
20.67 (0.01) & 0.41 (0.02) & 1.070 (0.009) & 0.39  (0.04)  & possibly enstatite + quartz? \\
21.65 (0.03) & 1.03 (0.06) & 1.093 (0.014) & 1.37  (0.30)  & possibly enstatite \\
23.09 (0.10) & 1.31 (0.51) & 1.189 (0.012) & 3.7   (1.7)   & unidentified \\
23.71 (0.02) & 0.58 (0.09) & 1.157 (0.028) & 1.4   (0.5)   & forsterite \\
27.52 (0.07) & 0.38 (0.12) & 1.070 (0.032) & 0.47  (0.26)  & forsterite \\
28.01 (0.04) & 0.66 (0.17) & 1.139 (0.032) & 1.5   (0.4)   & enstatite \\
29.08 (0.04) & 0.66 (0.09) & 1.079 (0.011) & 0.91  (0.17)  & enstatite \\
29.74 (0.03) & 0.66 (0.05) & 1.199 (0.025) & 2.2   (0.4)   & unidentified \\
30.51 (0.01) & 0.44 (0.04) & 1.165 (0.014) & 1.23  (0.12)  & unidentified \\
31.11 (0.02) & 0.24 (0.03) & 1.050 (0.004) & 0.209 (0.28)  & unidentified \\
32.73 (0.02) & 0.94 (0.12) & 1.236 (0.018) & 3.6   (0.5)   & unidentified \\
33.79 (0.01) & 0.98 (0.04) & 1.441 (0.030) & 6.8   (0.5)   & forsterite \\
39.86 (0.06) & 1.25 (0.14) & 1.092 (0.009) & 1.49  (0.05)  & unidentified \\
40.49 (0.05) & 0.80 (0.03) & 1.176 (0.014) & 1.82  (0.20)  & enstatite \\
41.56 (0.06) & 0.92 (0.21) & 1.085 (0.013) & 1.00  (0.34)  & unidentified \\
42.73 (0.09) & 0.89 (0.14) & 1.152 (0.035) & 1.71  (0.7)   & enstatite \\
43.54 (0.06) & 0.83 (0.21) & 1.179 (0.023) & 1.8   (0.6)   & crystalline H$_2$O \\
44.58 (0.01) & 0.78 (0.28) & 1.130 (0.004) & 1.2   (0.5)   & unidentified \\
47.65 (0.03) & 0.85 (0.08) & 1.107 (0.010) & 0.99  (0.06)  & unidentified \\
48.50 (0.01) & 0.55 (0.02) & 1.053 (0.004) & 0.312 (0.016) & unidentified \\
52.04 (0.13) & 3.8  (1.0)  & 1.036 (0.006) & 1.4   (0.6)   & crystalline H$_2$O/silicate?\\
63.56 (0.26) & 9.86 (0.33) & 1.110 (0.010) & 7.8   (0.9)   & crystalline H$_2$O + diopside + enstatite\\
68.96 (0.01) & 0.64 (0.02) & 1.069 (0.004) & 0.264 (0.008) & forsterite \\
91.12 (0.05) & 12.7 (0.7)  & 1.054 (0.006) & 1.87  (0.29)  & unidentified \\
\hline
\end{tabular}}
\label{tab:feat1}
\end{table*}

To aid the identification of the spectral features, we have subtracted a spline
curve `continuum' fit to the spectrum. The fit was made through selected points
where no feature was believed to be present, with the only other two 
constraints being that the fit was smooth (both in $F_{\nu}$ as 
well as in $F_{\lambda}$) and contained as much continuum as possible. 
This approach will however reduce the strength
of the features; it is also possible that very broad  features have been
treated as continuum. Note that a spline fit continuum has no physical meaning
and is only used to enhance the visibility of the sharp solid state features.
The continuum-subtracted spectrum is shown in Fig.~\ref{fig:contsub}. We find
a wealth of solid state features. The dust features naturally divide the
spectrum into a carbon-rich and an oxygen-rich part. Apart from the forbidden
emission lines, the spectrum below 15~$\mu$m is dominated
by the aromatic infrared bands (AIBs) and above 15~$\mu$m by O-rich dust
features.

In Table~\ref{tab:feat1} we list features found in the spectrum. The features
were fitted with Gaussians using a `local' continuum, i.e. the flux apart 
from the feature(s), and which was not necessarily
the same as the overall continuum. Since the errors are largely determined by
this local continuum, we estimated our errors by fitting the data for different
baseline continua. The spread in these results were used to calculate the
errors, which are quoted in Table~\ref{tab:feat1}. This local fitting may
prevent the discernment of some broad spectral features under narrow
features. Although most features are fitted reasonably well with a Gaussian
shape, we do not claim this to be the real shape. Since a lot of features
are blended with others, this method is
a systematical and reproducable way of determining the strength and position 
of the features.
Features visible at 12.8, 13.6, 14.2, 14.8, 15.3, 15.8, 16.2 and
25.8~$\mu$m are not listed since they are possibly instrumental artifacts,
caused by the presence of very strong forbidden emission lines 
on top of them (see below).

In the following sub-sections, we identify the different structures found
in the continuum-subtracted spectrum. In case of doubt about the
reality of features in the SWS wavelength range we compared the
two scan directions and, if possible, the AOT06 spectra of NGC~6302
taken on 20 February and 9 March 1997, with the final spectrum. For the LWS
range we checked for the presence of the feature in the individual 
observations, scans and sub-spectra.

\subsection{The 2.4 -- 12 $\mu$m region; C-rich dust}

The continuum subtracted spectrum from 2.4 to 12~$\mu$m (Fig. 2) 
is dominated
by the well known AIBs at 3.3, 3.4, 6.2, 7.7, 8.5, 11.0 and 11.3~$\mu$m, 
usually attributed to PAHs (Allamandola et al. 1989, Puget \& Leger 1989). 
Hence there is a clear indication of a C-rich dust environment, 
as first found by Roche \& Aitken (1986).  
A plateau is present at the long
wavelength side of the 11.3~$\mu$m feature extending to roughly  11.7~$\mu$m.

\subsection{The 12 -- 17 $\mu$m region; An uncertain area of the spectrum}

The reality and strength of the features from 12 to 17~$\mu$m is uncertain.
We do see structures in our spectrum at 12.8, 13.6, 14.2, 14.8, 15.3, 15.8,
and 16.2~$\mu$m. In other sources PAH features are seen at 12.8, 13.6,
14.2 and 16.2~$\mu$m (Hony et al. 2000; Van Kerckhoven et al. 2000).
However this region in NGC~6302 is one with many very strong
forbidden emission lines, some of which lie on top of the continuum features.
The resultant shape of the continuum may be perturbed due to detector problems
caused by shotnoise (Morris, private communication) and a comparison between
the up and down scans does reveal that the two scan directions show a
systematic offset with respect to each other at these wavelengths. We have also
checked the SWS-AOT06 observation and found similar differences between the up
and down scans, but also the absence of e.g. the 16.2~$\mu$m feature. 
Further, there is also evidence for spurious features in the
instrumental responsivity files at 13.6 and 14.2 $\mu$m (Morris, private
communication). Therefore although features are likely to be present in
this wavelength region, the reality of the shape, strength and  
even presence of the features found in the 12--17~$\mu$m
spectrum of NGC~6302 remains uncertain, so they are not listed in
Table~\ref{tab:feat1}.

\subsection{The 17 -- 200 $\mu$m region; O-rich dust}

The spectral region between 15 and 50 $\mu$m is characterized by a wealth of
relatively sharp emission features. Most of the emission features can be
identified with the Mg-rich crystalline silicates: forsterite (Mg$_2$SiO$_4$)
and enstatite (MgSiO$_3$). The identification of the crystalline silicate
features is based on the laboratory measurements of Koike \& Shibai (1998) and
J\"{a}ger et al. (1998), see Table~\ref{tab:feat1}. The wavelength position
of the different features, especially at the long wavelengths, indicate
that there is hardly any Fe present (Mg/(Fe+Mg) $>$ 0.99)
in the crystalline silicates.
The different laboratory data sets for enstatite, which in 
general are in agreement about the position and strength of the 
features, do not agree in the details of their spectral shape.
Therefore it is difficult to firmly identify some of these
features to enstatite and in Table~\ref{tab:feat1} they are listed as 
`possibly enstatite'. 

Amorphous silicates may well be present as indicated by the relatively broad
emission feature at 18~$\mu$m  (not separately measured in
Table~\ref{tab:feat1}), which is too broad to be just due to crystalline
silicates. The presence of a 10~$\mu$m silicate feature is suppressed by the
low temperature of the (silicate) dust (see Sect.~\ref{sec:model})
and the C-rich dust continuum which is
dominant at that wavelength. Another point which favours the presence of an
amorphous dust component besides the crystalline dust, is the relative weakness
of the crystalline dust features in the spectrum of NGC~6302 in this 
wavelength range
compared with the laboratory measurements of pure crystalline silicates.
Other sources with crystalline silicates also show evidence for
amorphous silicates e.g. AFGL 4106 (Molster et al. 1999b). It is of course
possible that other materials make a contribution to the 18~micron feature
and this would give an alternative explanation for the
suppression of the 10~$\mu$m band, but we have found no reasonable candidate
(yet).

The strength of the 20.7 $\mu$m band is not completely reproduced by enstatite
(see Sect.~\ref{sec:model}). This might be an indication of the presence of
quartz (SiO$_2$), which shows a very prominent peak at this wavelength.
Unfortunately, the other emission peaks of quartz are at much shorter
wavelengths, and are suppressed due to the quartz being at low temperature
compared to the hot C-rich dust.
Therefore it is difficult to prove the presence of quartz
based on this single feature. If quartz is present it will be only in small
amounts.

The broad feature peaking at 63.6 $\mu$m, hereafter referred to as the
60~$\mu$m feature, is probably a blend of diopside (Koike et al. 2000) and
crystalline H$_2$O ice with probably a small contribution from enstatite.
This feature, as well as the 43-$\mu$m ice feature, was observed by 
Omont et al. (1990) with KAO in the spectrum of IRAS~09371+1212 (Frosty Leo), 
while Barlow (1998)
reported the presence of 60 $\mu$m emission in several dusty circumstellar
shells and Sylvester et al. (1999) reported the ice emission in several
OH/IR stars. 
Laboratory data published by Bertie et al. (1969) and Schmitt et al.
(1998) show that
crystalline H$_2$O ice has two strong bands, one at about 43 $\mu$m, the other
at roughly 62~$\mu$m with a weak shoulder at 52~$\mu$m. 
All these features are also present in the spectrum of NGC~6302 and especially
the 43 and 60 micron feature cannot be explained by enstatite and/or diopside 
alone. Together with the expected presence of water-ice in a cold O-rich
environment, the identification of crystalline water ice is very robust.
However, it should be noted that a 52~$\mu$m feature is also seen in 
$\eta$~Car, which has no further indication for crystalline H$_2$O ice.
Therefore we do not exclude that this feature is (a blend with) 
another dust species, probably a crystalline silicate.

There remain a handful of yet unidentified features, indicating the presence of
other dust species. The broad feature at 91~$\mu$m indicates that there must be
a very cool dust component present. A feature around 100~$\mu$m in
HD147527 was suggested by Malfait et al. (1999) to be due to the presence of a
hydrous silicate. Hydrous silicates often show a broad peak
in this wavelength range (Koike \& Shibai 1990) and should only exist at low
temperatures. However, we do not know of a mechanism that would
produce hydrous silicates in the surroundings of NGC~6302. Furthermore
the peak in HD147527 is quite shifted from the one in NGC~6302, which makes
it somewhat unlikely that this is the same material. 
As very little data for silicate minerals exists in this wavelength region, 
the identification remains open.

\section{Fitting the features and continuum}
\label{sec:model}

Assuming that the dust shell is optically thin for wavelengths above 20 $\mu$m,
we applied a very simple procedure to get an estimate for the relative amount
of the crystalline and amorphous silicate dust fraction in the circumstellar 
dust shell around NGC~6302.

\begin{figure}[ht]
\centerline{\psfig{figure=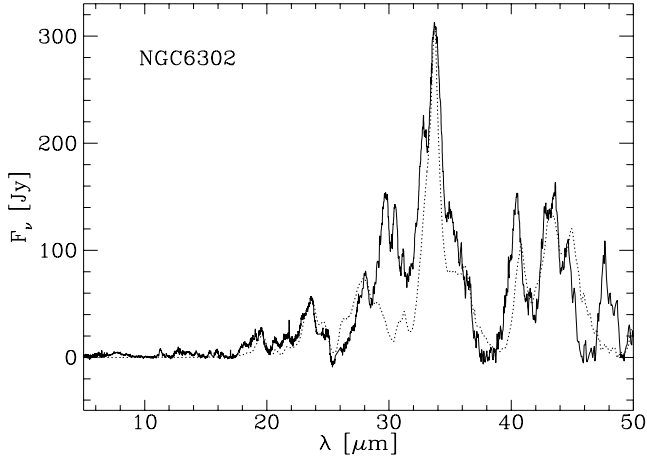,width=85mm,angle=270}}
\caption[]{\small A fit (dotted line) to the continuum subtracted spectrum 
(solid line) of NGC6302.
The mass absorption coeffcients of forsterite and the enstatite mix have 
been multiplied with Planck curves of 65 and 70~K, respectively.}
\label{fig:fit}
\end{figure}

In order to determine the temperature and relative abundance of enstatite
and forsterite we fitted continuum subtracted laboratory spectra to the 
continuum subtracted ISO spectrum of NGC~6302. 
Since one of the requirements of the spline fit to the continuum of NGC~6302 
was that it should contain as much flux as possible, 
it also contains continuum flux caused by the crystalline silicate components.
Although this is a small contribution, it is not negligible.
So, we subtracted also a continuum of the laboratory
spectra in the same way as we did for NGC~6302, i.e. by removing as much
flux as possible but keeping the continuum smooth. 
Fitting the continuum subtracted spectrum has the advantage that one can
immediately fit the temperature. It is also insensitive to the
temperature of the continuum, which might be different from the
temperature of the dust features. 

The temperature of forsterite is mainly determined by the
strength ratio of the features at 23.71 and 33.79~$\mu$m, while the
temperature of enstatite depends on the ratio of the features around 28 and 
40~$\mu$m. The laboratory data of Koike et al. (1999) were taken to fit the
spectra.
In this way we were able to determine independently 
the temperature of forsterite (65 $\pm$ 10K) and of enstatite (70 $\pm$ 10K),
see Fig.~\ref{fig:fit}.
The main uncertainty in the derived temperature is caused by the determination
of the continuum subtraction.
Based on the laboratory mass absorption coefficients we were able to derive
relative abundances of forsterite and enstatite. The derived
enstatite over forsterite mass ratio is 1.0, i.e. they are found in equal 
amounts. The typical error in the mass ratios is
a factor 2. This is mainly due to uncertainties in the temperature, which 
directly translate into differences in the mass fraction necessary to explain 
the strength of the features.

\begin{figure}[ht]
\centerline{\psfig{figure=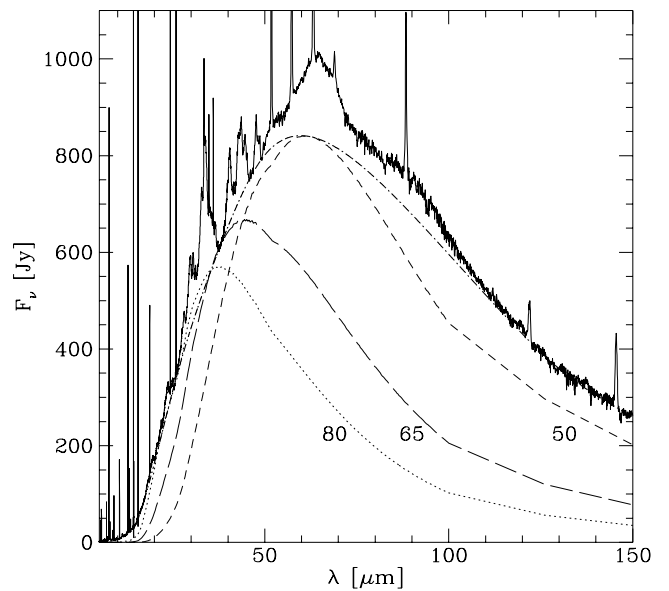,width=85mm,angle=0}}
\caption[]{\small The different fits to the spectrum of NGC6302 (solid line).
The dashed-dotted line is the continuum.
The astronomical silicate is represented at 3 different temperatures
at 80~K (dotted), 65~K (long-dashed) and 50~K (short-dashed).
The kinks after 90~$\mu$m are due to bad spectral coverage.}
\label{fig:amorf}
\end{figure}

The temperature and relative abundance of the amorphous silicates have been
determined by assuming that this material is mainly responsible for the
infrared excess. We fitted a single temperature single grain (0.1 $\mu$m)
astronomical silicate, data 
from Ossenkopf et al. (1992), to the complete (not continuum subtracted) 
spectrum. Part of the continuum
contains a small contribution of the crystalline silicate dust, the relative
mass estimate of the amorphous silicate will be slightly 
(not more than a few percent) overestimated, 
but this is negligible compared to the other sources of uncertainty.
The first major uncertainty is the fact that the infrared excess is 
too broad for a single temperature, single grain size approach. 
A fit to the spectrum up to 35~$\mu$m results in a temperature of about 80~K, 
but in a clear underestimation of the flux at longer wavelengths 
(see Fig.~\ref{fig:amorf}).
A fit which would contain most of the flux results in a temperature of
about 50~K (see Fig.~\ref{fig:amorf}). However, it is clear that this fit underestimates the flux
at these wavelengths were the temperature and mass of the crystalline 
silicates has been determined ($< 45 \mu$m). It also is a much lower 
temperature than what has been derived for forsterite and seems therefore 
unlikely.
Finally we also tried a fit with a similar temperature as found for forsterite
(65~K, see Fig.~\ref{fig:amorf}).
The amorphous silicate over forsterite mass ratios derived for these
different temperatures are 6.3, 20.4 and 97 for respectively 80, 65 and 50~K.

The gentle slope up to the 1.1mm point (Hoare et al. 1992) indicates that
large grains ($>10~\mu$m) are present. They will be responsible for 
most of the flux at the long wavelengths, but due to their low temperature
not for the flux below 40~$\mu$m. We therefore assume that the typical
temperature of the small amorphous silicates, which are the source of flux 
below 40~$\mu$m, is somewhere between 65 and 80~K (see Fig~\ref{fig:amorf}).
We have calculated the dust mass necessary to explain the strength of the
curves in Fig~\ref{fig:amorf} assuming a grainsize of 0.1 $\mu$m. 
We found that the fraction of crystalline silicates 
(enstatite and forsterite) in the total mass of small dust grains will be 
between 9 (when the small amorphous silicate dust grains are 65~K) and 24~\% 
(when they are 80~K).
The composition and thus the abundance of the large grains is very difficult to
determine, since they will not show spectral features.

We have made no effort to calculate the abundance of the water-ice or diopside.
Both contribute significantly to the feature at 60~$\mu$m. Because of 
their blending it is very difficult to disentangle them and derive estimates 
for their temperature and thus their mass fraction.

As shown above the temperature for the amorphous silicates is difficult to 
determine, and the width of the infrared excess suggests a temperature 
distribution. The slope below 40~$\mu$m suggests the presence of
a population of relative warm (i.e. warmer than forsterite and enstatite)
amorphous grains.
We would like to note that in other studies, with
full radiative transfer modeling (Molster et al. 1999b; 2001), it is shown that
amorphous silicates are warmer than crystalline silicates. This may be due
to a difference 
in chemical composition (i.e. the lack of Fe in the crystalline silicates), 
which results in different absorptivities in the UV, visual and 
near-IR wavelengths for the crystalline and amorphous silicates.

\section{Discussion}
\label{discussion}

Both C-rich (PAH features) and O-rich (silicate features) chemistry is present
in the spectrum of NGC~6302.
A mixed chemistry is remarkable in an outflow, since one of the first formed 
stable molecules in an outflow is CO. The formation of CO is very efficient
and completely 'absorbs' the less abundant of the two. 
This leaves the most abundant atom to determine the dust composition.
In the context of the massive N-rich Type~I planetary nebula NGC~6302,
Justtanont et al.~(1992) and Barlow~(1993)
pointed out that hot-bottom burning (HBB) at the base of the hydrogen
envelopes of
sufficiently massive AGB stars can eventually convert carbon dredged-up after a
thermal pulse to nitrogen via the CN cycle. Thus the surface C/O ratios and
the grains condensed, can go from O-rich to C-rich and back again after each
of the multiple thermal pulses experienced. This would allow both types of grains to be
present in the resulting nebulae, without a single transition radius from
O-rich to C-rich material. This scenario would produce a mixed chemistry
in the whole nebula.

It is difficult in this instance to explain the strength of the crystalline 
silicate features. They are much stronger than what is seen in normal
outflow sources (Molster et al. 1999a).
However, there are other evolved stars which show PAHs and crystalline
silicates in their spectrum and are supposed to follow a different scenario.
A famous example of this mixed chemistry is the
Red Rectangle (Waters et al. 1998). This binary system formed a stable
circumbinary disk while the star was still oxygen-rich.
While the star evolved to a carbon-rich star, the disk remained.
This is evidenced by the carbon-rich outflow and the oxygen-rich disk.
The presence of a toroidal disk in NGC~6302 has long been
known (Meaburn \& Walsh 1980;
Lester \& Dinerstein 1984; Rodriguez et al. 1985),
which, if stable, can explain the high degree of crystallinity by processes
suggested  by Molster et al. (1999a). The relatively high mm flux of NGC~6302
(Hoare et al. 1992)
points to the presence of a population of large and cold grains.
The presence of a stable disk naturally provides a suitable 
surrounding to produce large, and therefore cold, grains.  
This scenario would predict the O-rich material only to be present in the
disk, with the C-rich dust dominating in the outflow and less
pronounced in the disk.

In this context it is interesting to
compare NGC~6302 with CPD$-56^{\circ}8032$. Although the central stars of
these PNe are different, the spectra of the dust shells look quite similar.
Both have roughly the same degree of crystallization, and the crystalline
silicates have similar temperatures  (Cohen et al. 1999). This is quite
striking since the nebula around CPD$-56^{\circ}8032$ is expected to be much
younger ($\approx 10^2$ yr, De Marco et al. 1997) than the nebula  around
NGC~6302 ($\approx 10^4$ yr, Terzian 1997). The expansion velocities are  not
that different, therefore it is likely that the crystalline material  is not
formed in the present day outflow, but more likely was already present
around these
objects, e.g. in a disk, before this outflow started. It has been suggested
that we see here the destruction of Oort-cloud like objects (Cohen et al.
1999), however the degree of crystallization is much higher than the
crystallinity so far known for comets and 
interplanetary dust particles in our own solar
system, which may make this Oort-cloud scenario unlikely.

The weakness of the PAH features with respect to the rest of the spectrum
in NGC~6302 resembles the strength of the PAH features in the spectra of
Luminous Blue Variables (LBVs) like R71 and AG Car, which show weak PAH
emission and relatively strong crystalline silicates
(Voors et al. 1999; Voors et al. 2000).
Since these massive stars have a C/O ratio lower than 1, the
formation of PAHs is not understood. It has been suggested that
shocks might dissociate the CO and that the PAHs form in the after-shock
region. If the PAHs in NGC~6302 are formed in a similar way, one would expect
to find the PAHs in the neighbourhood of these shocked regions.
Casassus et al. (2000) show that the 3.3 $\mu$m emission
is predominantly coming from the central region where the present-day outflow
is expected to collide with the disk. Also the C/O ratio of 0.88 (see Casassus
et al. 2000) points to an oxygen-rich outflow. In the case of this scenario
it is expected that the outflow and disk will be oxygen-rich and PAHs are only
seen in the (after-)shocked regions. This would even allow the star to have 
been oxygen-rich for its entire life.

This last scenario explains the observed features best.
However, one should keep in
mind that a thorough  understanding of the PAH formation in an oxygen
dominated environment is still lacking.
All three scenarios predict a different dust distribution.
With the aid of future high spatial resolution observations 
in the PAH and crystalline silicate
bands (and the continuum) we may be able
to distinguish between the HBB scenario and the two different
disk scenarios for NGC~6302.

\acknowledgements{FJM acknowledges support from NWO under
grant 781-71-052 and under the talent fellowship program.
LBFMW acknowledge financial support from an NWO 'Pionier' grant.
This work was partly supported by NWO Spinoza grant 08-0 to
E.P.J. van den Heuvel. MC acknowledges
the support of NASA under grant NAG 5-4884 and contract 961501
through JPL.}

\end{document}